\documentclass[aps,eqsecnum,twocolumn,nofootinbib,prd]{revtex4}
\usepackage{amsmath}
\usepackage{amssymb}
\usepackage{amsbsy}
\usepackage{epsfig}
\usepackage{bm}

\begin{document}

\title{\Large On the study of neutrino properties}
\author{Y. H. Yuan}
\email[E-mail: ]{henry@physics.wisc.edu}
\date{December 12, 2005}

\begin{abstract}

We review the discovery of the neutrino and outline the history 
of neutrino physics. Many interesting phenomena involving the neutrino are exhibited. 
We also discuss the long-standing solar neutrino puzzle and the properties 
of the neutrino which lead to various important results. We present a possible experimental test of the neutrino property. In addition, neutrino oscillation and neutrino spin precession are also demonstrated. 

\end{abstract}

\maketitle

\section{Introduction}

The elusive neutrino\footnote{In this paper, neutrino means electron neutrino only except when specified.} has played an important role in our understanding
of physics in many ways: from the parity violation\cite{ly,wu,ga} of
beta decay to the solar neutrino puzzle; from the formulation of the
four Fermion weak interaction theory to the unification of 
electromagnetic and weak interactions into the electroweak
interaction\cite{gl,sa,we}. There are various open questions about neutrinos that need both 
theoretical and experimental exploration. 
As the most intriguing and fascinating fundamental particle, the neutrino is so important that neutrino physics has become one of the most
significant branches of modern physics. 

On December 4, 1930, W. Pauli proposed the neutrino as a
desperate-remedy to the observed continuous spectra of energy for the
outgoing electrons of beta decay. In accurate measurements on beta 
decay
process before 1930, physicists found the emited electron with a
continuous energy spectrum, unlike alpha decay and gamma decay in which
the emitted particles carried away the well-defined energy which is
equal to the total energy difference of the initial and final states. 
It meant that a particular nucleus emitted an electron bearing
unpredictable energy in a particular transition. This experimental
result apparently violated the conservation laws of energy and 
momentum.
In order to solve this serious problem Pauli proposed an entirely new
particle in his open letter\cite{pa} to the group of radioactives at 
the
meeting 
of the regional society in Tubingen:

``...This is the possibility that there might exist in the nuclei
electrically 
neutral particles, which I shall call 
neutrons, which have spin 1/2, obey the exclusion principle and 
moreover differ from light quanta in not travelling with the velocity 
of light.''

``... I admit that my remedy may perhaps appear unlikely from 
the start, since one probably would long ago have seen the neutrons if 
they existed. But `nothing venture, nothing win', and the gravity of 
the situation with regard to the continuous beta spectrum is 
illuminated by a pronouncement of my respected predecessor in office, 
Herr Debye, who recently said to me in Brussels `Oh, it's best not to 
think about it at all-like the new taxes'. One ought to discuss
seriously every avenue of rescue."

In his letter, Pauli called his new proposed particle-the "neutron"
which is now called neutrino due to Enrico Fermi. Pauli proposed that
this new speculative neutral particle might resolve the nonconservation
of energy. If the proposed neutrino and the electron were emitted
simultaneously, the continuous spectum of energy might be explained by
the sharing of energy and momentum of emitted particles in beta decay.
The neutrino was first experimentally detected by Fred Reines and Clyde
Cowan in 1956\cite{re} using a liquid scintillation device. Their
experiment involved detecting the reaction $p+\nu_e\to n+e^+$ 
exploiting
antineutrinos from the Savannah river nuclear reactor. This important
discovery won the 1995 Nobel prize in physics. It is worth mentioning
that long before the neutrino was experimentally detected, Enrico
Fermi\cite{fe} incorporated Pauli's proposal in his brilliant model for
beta decay in the framework of quantum electrodynamics in 1934. He
showed clearly 
with his beta decay theory that the neutron decayed into a proton, an
electron and a neutrino simultaneously. In 1957, B. Pontecorvo\cite{pv}
suggested that neutrino flavor eigenstates are superpositions of its
mass eigenstates, thus as the neutrino propogate it would undergo
oscillation, which is just similar to the $K$ meson system. The little
neutrino has found its application to a number of different research
areas in physics, such as in particle physics, nuclear physics,
cosmology and astrophysics. Thanks to the conjecture of the
neutrino by Pauli which rescues the
fundamental conservation laws of energy and momentum. Although his proposal contradicted the well-accepted knowledge at the time on beta decay process, his new beta decay process
involving the neutrino was not completely impossible experimentally.

\section{The properties of the neutrino}

The solar neutrino problem is a famous puzzle on the neutrino. The sun
shines mainly because of the hydrogen burning. The nuclear fusion
reaction may be written as,
\begin{center}
\begin{equation}
4p \to ^4He + 2e^+ + 2 \nu_e 
\end{equation}
\end{center}

The positions produced in the above nuclear fusion reaction was
annihilated with electrons while the emitted neutrinos hardly and weakly interact with matter therefore the sun may be regarded as a well-defined neutrino source shown in FIG.1. Thanks to the sun. So we may have the
great opportunity to study the properties of the neutrino. The
pioneering work of detecting the solar neutrino was carried out by R.
Davis\cite{da}. His Homestake chlorine experiment was based upon the
following reaction,

\begin{equation}
 ^{37}Cl + \nu_e  \to  ^{37}Ar + e^- 
\end{equation}

When detecting the neutrino emitted from the sun, Davis consistently observed fewer solar neutrino
capture rate than the calculated capture rate predicted by J. Bahcall\cite{ba} in
chlorine using detailed computer models of the solar interior in 1968. Later the missing neutrino mystery was also observed by other groups using different materials. This is the long-standing solar
neutrino puzzel. 

The SNO experiment \cite{sn} detected the solar neutrino which showed the
flavor change of the neutrino. They measured only the high energy $^8B$
solar neutrinos through the reactions,

\begin{center}
  \begin{tabular}{ll}
     $\nu_e + d \rightarrow p + p + e^-$\hspace{0.5in} & (CC)\\
     $ \nu_x + d \rightarrow p + n + \nu_x$ & (NC)\\
     $ \nu_x + e^- \rightarrow \nu_x + e^-$  & (ES)\\        
  \end{tabular}
 \end{center}.

In their measurement of the $^8$B neutrino fluxes, they assumed the
standard spectrum shape\cite{ce} and obtained,

\begin{eqnarray*}
\phi_{\text{SNO}}^{\text{CC}}(\nu_e) & = & 1.75 \pm 0.07~({\rm stat.})
      ^{+0.12}_{-0.11}~({\rm sys.}) \pm 0.05~({\rm theor.}) \\
                              &   & \times 10^6~{\rm cm}^{-2} {\rm
s}^{-1} \\
\phi_{\text{SNO}}^{\text{ES}}(\nu_x) & = & 2.39 \pm 0.34 ({\rm stat.})
      ^{+0.16}_{-0.14}~({\rm sys.}) \times 10^6~{\rm cm}^{-2} {\rm s}^
{-1}
\end{eqnarray*}

Their total flux of active $^8$B neutrinos is, 
\begin{equation}
\phi(\nu_x) =  5.44\pm 0.99 \times 10^6~{\rm cm}^{-2}
{\rm s}^{-1}. 
\end{equation}
which agrees with the predictions made from the standard solar
models\cite{ba,TC}.

Although SNO uses heavy water to detect the neutrino, as far as the
charge current reaction is concerned, SNO experiments are very similar
to the ones carried out at Super-Kamiokande\cite{sk,ya}, which also show
that the neutrino flavor change. They are both real time
experiments sensitive to $^8B$ solar neutrinos only with the Cherenkov
detector. When comparing their measured flux via charge current 
reaction
with the flux obtained by Super-Kamiokande Collaboration of the $^8$B
flux using the elastic scattering reaction, they found the difference
was $0.57\pm 0.17 \times 10^{6}$ cm$^{-2}{\rm s}^{-1}$, which was about
3.3$\sigma$. This gives the direct evidence of the flavor change of the
neutrino. As a consequence of the flavor change, the neutrinos should
have mass. The neutrino flavor change was also justified by the KamLAND
reactor neutrino experiment with liquid scintillator detector located 
at the old Kamiokande site\cite{kl}. 

A great number of explanations were proposed to solve the solar neutrino puzzle.
The most popular one is the neutrino oscillation. Next we 
will show the three flavor neutrino oscillation with the plane wave
approximation. A definite neutrino flavor field $\nu_f$ with flavor $f$ is a linear
combination of the neutrino mass fields $\nu_m$ with the definite mass $M_m$ and
definite energy $E_m$, so

\begin{equation}
\nu_f= \sum^{3}_{m=1} U_{fm}\nu_m
\end{equation}
where$ f = e,\mu,\tau$, $U_{fm}$ are the entries of a unitary matrix.
Then we obtain,

\begin{equation}
|\nu_{f}\rangle =\sum^{3}_{m=1}U^{*}_{fm} |\nu_{m}\rangle 
\label{mm}
\end{equation}
namely the neutrino flavor state is the superposition of the neutrino mass
eigenstates. Next, considering the neutrino mass state at time t, $|\nu_{m}\rangle_t$, using the
Schrodinger equation yields,

\begin{equation}
|\nu_m\rangle_t =e^{-iE_m t} |\nu_{m}\rangle
\end{equation}
where $|\nu_{m}\rangle$ represents the neutrino mass state at time 0 in
its rest frame. Combining with Eq.(\ref{mm}) we immediately obtain the
neutrino flavor state at time t,

\begin{eqnarray}
|\nu_{f}\rangle_t 
& =&\sum^{3}_{m=1}U^{*}_{fm} e^{-iE_{m}t}|\nu_{m}\rangle   \nonumber  
\\
&=& \sum_{f'}\left(\sum^3_{m=1} U_{fm}^* e^{-iE_m t}U_{f'm}
\right)|\nu_{f'} \rangle  \nonumber  \\
&=&  \sum_{f'}M(\nu_f\to\nu_{f'})|\nu_{f'} \rangle
\label{nn}
\end{eqnarray}
where $f'=e,\mu,\tau$ and $M(\nu_f\to\nu_{f'})$ represents the amplitude of
neutrino flavor transition at time t. That the neutrino initially with flavor 
f turns into a superposition of different neutrino flavors after 
traveling time t is clearly shown in Eq.(\ref{nn}). Therefore we immediately obtain 
the probability of the neutrino flavor transition in vacuum from $\nu_f$ to $\nu_{f'}$,

\begin{eqnarray}
P &=& \left| M(\nu_f\to\nu_{f'}) \right|^2   \nonumber  \\
&=& \left| \sum^3_{m=1} U_{fm}^* \, e^{- i E_m t} \, U_{f'm} \right|^2 
\nonumber  \\
&=&\sum_{m,m'} U_{fm}^*U_{f'm}U_{fm'}U_{f'm'}^*e^{-i( E_m - E_{m'})t}
\label{pp}
\end{eqnarray}
where $m,m'=1,2,3$ representing the neutrino mass eigenstates. Next
exploit the mass-energy relation due to Einstein,

\begin{equation}
E_m=\sqrt{p^2_m+m_m^2}\approx p+\frac{m_m^2}{2p}
\label{qq}
\end{equation}
where we have assumed that all neutrino mass eigenstates have the same momentum $p$ and employed the relation $p\gg m_m $ because of the high energy neutrinos observed.  When plugging Eq.(\ref{qq}) into Eq.(\ref{pp}), yields the probability of the neutrino flavor transition in vacuum from $ \nu_f$ to $\nu_{f'}$,

\begin{eqnarray}
P&=&\sum_{m,m'}
U_{fm}^*U_{f'm}U_{fm'}U_{f'm'}^*e^{-i(\frac{m_m^2 - m_{m'}^2}{2p})t} \nonumber  \\
&=&U_{fm}^*U_{f'm}U_{fm'}U_{f'm'}^*e^{-i(\frac{\Delta{m}^2_{mm'}t}{2E})t} 
\end{eqnarray}
where $E$ is the neutrino energy in the massless limit which is approximately equal to its momentum $p$ due to the extremely small mass of the neutrino and we have defined,

\begin{equation}
\Delta{m}^2_{mm'} \equiv m_m^2 - m_{m'}^2
\end{equation}.

Another possible way in explaining the solar neutrino puzzle is
spin-flavor precession\cite{ac,vv,ok}. It is based upon the neutrino spin
precession through the strong magnetic field in the convective 
zone of the sun. Following their proposals, the neutrino has a magnetic
moment $\mu \approx (10^{-11 }-10^{-10})\mu_B$ where Bohr magneton
$\mu_B=\frac{e\hbar}{2m_e c}$. The spin of the neutrino 
would precess from a left-handed to a right-handed helicity due to the magnetic moment as well as the electric dipole moment the neutrino has when the neutrino passes through the strong interior solar magnetic field. To be more specific, a $\nu_{eL}$ may flip its spin and change its flavor then 
turn into $\nu_{\mu R}$ or $\nu_{\tau R}$ (where $L$ and $R$ represent
left-handed and right-handed helicity respectively.) which would not
interact with the materials used in the solar neutrino experiments. As 
a consequence, the neutrino spin precession leads to the measured
neutrino deficit in the solar neutrino puzzle. The measured neutrino 
deficit may also be explained by particle interactions. Put in 
other word, when a neutrino interacts with an antineutrino they may produce muon neutrino 
and muon antineutrino pair. The cross section of this interaction would
be greatly increased if the neutrino possesses any charge. We propose 
that the neutrino has a magnetic charge which has similarly puzzled physicists for a long time. This assumptiom is justified by another famous interaction in which the
neutrino takes part--beta decay. As we know space inversion will not
conserve in the beta decay process. Actually in all weak interactions involving neutrinos, 
parity violations always happen. This behavior of the neutrino is quite similar to the 
behavior of monopole under space inversion. As J. Jackson pointed out 
in his famous book \cite{ja} on Classical Electrodynamics:``... it is a 
necessary consequence of the existence of a particle with both 
electric and magnetic charges that space inversion and time reversal are no longer 
valid symmetries of the laws of physics. It is a fact, of course, that these 
symmetry principles are not exactly valid in the realm of elementary 
particle physics, but present evidence is that their violation is 
extremely small and associated somehow with the 
weak interaction." Since magnetic charge density is a pseudoscalar, the signs of a magnetic 
charge are opposite when observed from both the right-handed coordinate system and the 
left-handed coordinate system. This could result in 
the parity violation of the weak interaction involving the neutrino. The neutrino has the electric dipole moment $\vec {d}$. Since the orientation of the 
neutrino in its rest frame is characterized only by 
the orientation of the internal vector --- intrinsic angular momentum 
$\vec{J}$, $\vec {d}$ and $\vec{J}$ must be either in the 
same direction or in the opposite direction. We obtain $
\vec {d}= {L}\frac{{g}\vec{J}}{2m}$ where $g$ is the magnetic charge a particle has, 
$m$ is the mass of that particle and the parameter $L$ is called a Lande factor. 
Due to the electric dipole moment of neutrino, nonconservation of time reversal of weak 
interactions involving neutrinos takes place. Moreover $T$ violation induce the $CP$ violation because of the L$\ddot{u}$ders--Pauli theorem. The electric dipole moment makes the neutrino spin precession explanation more convincing.
\begin{figure} [ht]
\centering 
\epsfig{file=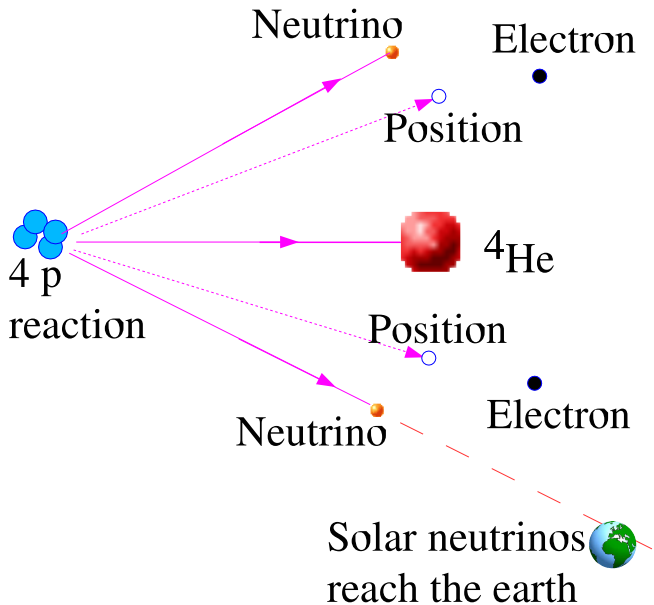,width=\columnwidth,angle=0} 
\caption{Diagram of the hydrogen burning in the sun which shows that the sun is a well-defined neutrino source. Through the nuclear fusion reaction of four hydrogen nuclei(namely protons), a helium nucleus, two positions and two neutrinos are produced. Then the positions are annihilated with electrons while the emitted neutrinos interact hardly and weakly. As a consequence, a number of solar neutrinos are observed on the earth.} 
\label{78}
\end{figure}
When the neutrino moving through the strong interior solar magnetic field, 
the magnetic field would cause neutrino spin precession, giving rise to the surprising discrepancy between the calculated and observed capture rates of solar neutrino. Before completing this section, 
we give a possible experimental test based upon the Faraday induction 
method. Place a radioactive source at the center of an enclosed superconducting 
sphere. Whenever $\beta$-decay of the radioactive source happens, an anti-neutrino is released 
which would induce the supercurrents on the superconducting sphere due 
to the proposal. A magnetometer SQUID (superconducting quantum 
interference device) connected to the sphere is need to monitor the 
currents. To eliminate any unwanted influence 
of electrically charged particles, an absorbent layer would be introduced between the 
radioactive source and the enclosed superconducting sphere. Even though, the sensitive devices in the experiment are vulnerable to spurious signals, it is still an ideal way to detect the monopole since the method is independence of the particle's mass and velocity. 
The detectors should be placed inside a magnetic shield made 
up of lead or mumetal to protect the detectors from external magnetic 
fields. 

\section{Summary and conclusion}

We have reviewed the discovery and the history of neutrino physics in our paper. A number of interesting phenomena involving the neutrino have been exhibited. 
We have studied the long-standing solar neutrino puzzle and
the properties of the neutrino which lead to various interesting 
results. In addition, neutrino oscillation and spin precession have also been discussed. We have presented a possible experimental test of the 
neutrino property. This year is the unprecedented World Year of Physics which marks the hundredth 
anniversary of the pioneering contributions of Albert Einstein. 
We dedicate this paper to Albert Einstein.


\begin{thebibliography}{10}
\bibitem{ly}
T. D. Lee and C. N. Yang, Phys. Rev. {\bf 104}, 254, (1956).
\bibitem{wu}
C. S. Wu, E. Ambler, R. W. Hayward, D. D. Hoppes and R. P. Hudson, 
Phys. Rev. {\bf 105}, 1413, (1957). 
\bibitem{ga}
R. L. Garwin, L. M. Lederman and M. Weinrich, Phys. Rev. {\bf 105}, 
\bibitem{gl}
S. L. Glashow, Nucl. Phys. 22, 579 (1961). 
\bibitem{sa}
A. Salam, in Elementary Particle Theory, Proc. of the 8th Nobel 
symposium, Aspenasgarden, 1968, ed. by N. Svartholm, p.367
\bibitem{we}
S. Weinberg, Phys. Rev. Lett. {\bf 19}, 1264 (1967).
\bibitem{pa}
Charles Enz, No time to be brief-A scientific biography of Wolfgang
Pauli, Oxford university press, (2002)
\bibitem{re}
F. Reines and C. Cowan, Science, 124 (1956) 10
\bibitem{fe}
E. Fermi, Z. fur Physik {\bf 88},161, (1934)
\bibitem{pv}
B. Pontecorvo, J. Exptl. Theoret. Phys. {\bf 33}, 549 (1957), Sov. Phys.
JETP, {\bf 6}, 429, (1957)
\bibitem{da}
R. Davis, Jr., D. S. Harmer and K. C. Hoffman, Phys. Rev. Lett {\bf 
20}, 1205, (1968).
\bibitem{ba} 
J.N. Bahcall, M. H. Pinsonneault, and S. Basu,
astro-ph/0010346 v2.
\bibitem{sn} 
SNO collaboration, 
 Q.R. Ahmed {\it et al.}, Phys. Rev. Lett. {\bf 87}, 071301 (2001); 
 Phys. Rev. Lett. {\bf 89}, 011301 (2002); 
 Phys. Rev. Lett. {\bf 89}, 011302 (2002) 
\bibitem{ce}
C. Ortiz {\it et al.}, Phys. Rev. Lett. {\bf 85}, 2909 (2000); 
\bibitem{TC} 
A.S. Brun, S. Turck-Chi\`{e}ze, and J.P. Zahn,
Astrophys. J. {\bf 525}, 1032 (1999);  S. Turck-Chi\`{e}ze {\em et 
al.}, 
Ap. J. Lett., v. {\bf 555} July 1, 2001. 
\bibitem{sk} 
Super-Kamiokande Collaboration, 
S. Fukuda {\it et al.}, Phys. Rev. Lett. {\bf 81}, 1562 (1998); 
S. Fukuda {\it et al.}, Phys. Rev. Lett. {\bf 86}, 5651 (2001).
\bibitem{ya} 
Y. Ashie {\it et al.}, Phys. Rev. Lett. {\bf 93}, 101801 (2004); 
Y. Ashie {\it et al.,}  Phys.Rev. D71 (2005) 112005.
\bibitem{kl} 
KamLAND collaboration, T.Araki {\it et al.}, Phys. Rev. Lett. {\bf 94},
081801 (2005). 
\bibitem{ac}
A. Cisneros, Astrophys. Space Sci.{\bf 10}, 87 (1971).
\bibitem{vv}
M.B. Voloshin and M.I. Vysotski, Yad. Fiz. (Sov. J. Nucl. Phys.) {\bf 44}, 845 (1986).
\bibitem{ok}
L.B. Okun, Yad. Fiz. (Sov. J. Nucl. Phys.) {\bf 44}, 847 (1986).
\bibitem{ja}
J. D. Jackson, Classical Electrodynamics, 2nd edition, John Wiley \&
Sons, New York, (1975)

\end{thebibliography}
\end{document}